# Low Effective Mass Leading to High Thermoelectric Performance

Yanzhong Pei, Aaron D. LaLonde, Heng Wang and G. Jeffrey Snyder

Materials Science, California Institute of Technology, Pasadena, CA 91125, USA.

High Seebeck coefficient by creating large density of state (DOS) around the Fermi level through either electronic structure modification or manipulating nanostructures, is commonly considered as a route to advanced thermoelectrics. However, large density of state due to flat bands leads to large effective mass, which results in a simultaneous decrease of mobility. In fact, the net effect of high effective mass is a lower thermoelectric figure of merit when the carriers are predominantly scattered by acoustic phonons according to the deformation potential theory of Bardeen-Shockley. We demonstrate the beneficial effect of light effective mass leading to high power factor in n-type thermoelectric PbTe, where doping and temperature can be used to tune the effective mass. This clear demonstration of the deformation potential theory to thermoelectrics shows that the guiding principle for band structure engineering should be low effective mass along the transport direction.

Increasing the thermoelectric figure of merit ($zT$) is the most challenging task to enable the widespread use of this method to directly convert heat into electricity. The transport properties including resistivity ($\rho$), Seebeck coefficient ($S$), electronic ($\kappa_E$) and lattice ($\kappa_L$) components of thermal conductivity ($\kappa=\kappa_E+\kappa_L$) determine the figure of merit, $zT=S^2T/\rho\kappa$, where $T$ is absolute temperature.

Creating phonon scattering centers such as nanostructures[1-4] to lower $\kappa_L$, has been proven effective for achieving $zT > 1$ in many instances. However, $\kappa_L$ in such materials already approaches their amorphous limit[3, 4], suggesting strategies targeting increases in $zT$ by improvements of the thermoelectric power factor ($S^2/\rho$).

The decoupling of $S$, $\rho$ and $\kappa_E$ in an effort to achieve high $zT$ has been a longstanding challenge as they are strongly coupled with each other through the carrier concentration, scattering and band structure[5-7]. However, it is well known that the optimal electronic performance of a thermoelectric semiconductor depends primarily on the weighted mobility[7-10], $\mu m^{*3/2}$, which includes both the density-of-states effective mass ($m^*$, with a unit of free electron mass $m_e$) and the nondegenerate mobility ($\mu$) of carriers.

More generally, each degenerate carrier pocket makes a contribution to $m^*$ via $m^*=N_v^{2/3}m_b^*$ [7-11], where $N_v$ and $m_b^*$ are the number of degenerate carrier pockets and the average band mass (density-of-states effective mass for each pocket), respectively. Without explicitly reducing $\mu$, converging many valence (or conduction) bands to achieve high $N_v$ and therefore a high $m^*$ has been proposed as an effective approach to high performance in both bulk[12] and low dimensional[13] thermoelectrics.

Without modifying $N_v$ and in an attempt to increase the power factor, many efforts have been recently devoted to increasing the Seebeck coefficient (i. e. increasing $m^*$ through high $m_b^*$) either by designing[14, 15] the density of states or manipulating nanostructures[16, 17]. This concept has recently been considered as a criteria[9, 10] for obtaining good thermoelectrics. However, these methods may reduce the mobility significantly[15].

In fact, an increase of $m^*$ resulting from increasing $m_b^*$ (i. e. by flattening the band), leads to a significant decrease in mobility according to the deformation potential theory of Bardeen-Shockley[18]. This is because $\mu \propto m_b^{*-3/2}m_c^{*-1}$ ($m_c^*$, conduction mass of the carriers) [18-20] when the carriers are predominantly scattered by acoustic phonons, as has been found in most of known and good thermoelectrics. As a result, the optimal power factor $S^2/\rho \propto \mu m^{*3/2} \propto N_v m_c^{*-1}$ becomes inversely proportional to $m_c^*$, the effective mass along the conducting direction[7, 8, 21, 22]. Since cubic thermoelectric materials such as PbTe, SiGe and skutterudites have an isotropic $m_c^*$ that increases with increasing $m_b^*$[23], it is then clear that increasing $m_b^*$ actually decreases the optimal power factor in spite of the resulting large Seebeck coefficient.

In this paper we demonstrate that a lower effective mass (either by doping or by adjusting the temperature) leads to a high power factor and thus excellent thermoelectric performance in n-PbTe. When compared with La-doped PbTe, a ~20% lower effective mass in I-doped PbTe results in a ~20% higher power factor. A single Kane band (SKB) model[21, 24, 25] has been developed to quantitatively understand that the lower effective mass is indeed beneficial for enhancing the thermoelectric performance. This work shows a contrasting example to the commonly utilized strategy for large Seebeck coefficient resulting from heavy band mass for high performance thermoelectrics[9, 14, 16, 17].

La- and I-doped PbTe ($La_xPb_{1-x}Te$ and $PbTe_{1-x}I_x$ with $0<x<0.01$) were synthesized by the same melting, quenching, annealing and hot pressing method. The synthesis procedure and details of the measurement of transport properties can be found elsewhere[26, 27]. In both La and I doped PbTe the donor states are very deep[28] so that each dopant atom produces one electron[27] in conduction band according to the rules of valence[29]. The Hall carrier concentration ($n_H=1/eR_H$, $e$ is electron charge) is determined from the measured Hall coefficient ($R_H$), and the room temperature values of $n_H$ were used to label the samples. All the samples in this study show n-type conduction. La- and I-doped PbTe samples with two important Hall carrier concentrations of ~1.8 and ~3×10$^{19}$ cm$^{-3}$, which respectively enable the highest average $zT$ and peak $zT$ in the temperature range of most interest for thermoelectric applications, were chosen for the discussion of temperature dependent transport properties. To avoid the detrimental effects due to minority carriers (Fig. 1) and to validate the use of the single band conduction model as discussed below, we focus on the transport properties from 300 to 600 K for this study. For temperatures or carrier concentration where the scattering is not dominated by acoustic (nonpolar) phonons or the transport properties are not sufficiently described by a single band model the following conclusions may differ.

The measured Seebeck coefficient, resistivity, thermal conductivity and $zT$ are shown in Fig. 1 as a function of temperature. The monotonically increasing Seebeck coefficient and resistivity, as well as the slightly (<10 %) increased Hall coefficient (which can be expected from a slight loss of degeneracy, not shown), with increasing temperature allows the assumption of single band conduction behavior at $T<\sim600$ K to be made in this study. This assumption is consistent with band structure studies of PbTe[21].

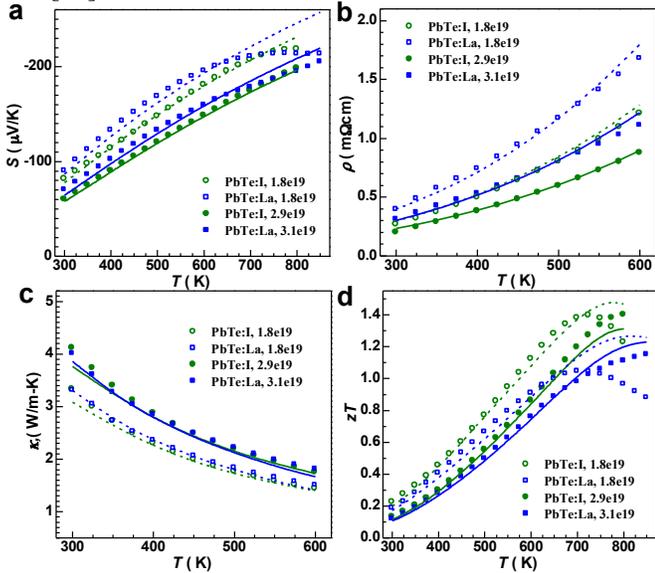

Fig. 1. Temperature dependent Seebeck coefficient (a), resisistivity (b), thermal conductivity (c) and thermoelectric figure of merit (d) for two groups of La- and I-doped PbTe having room temperature Hall carrier concentration of ~1.8 and ~3×10$^{19}$ cm$^{-3}$, respectively. The curves represent the predicted results from the single Kane band model with an effective mass of 0.25 $m_e$ for I-doping and 0.3 $m_e$ for La-doping, respectively. Comparing with La-doped PbTe that has nearly the same carrier concentration, a 20% lower effective mass in I-doped leads to ~20% higher figure of merit over the whole temperature considered.

It has been well known that the bands located at the L point ($N_v$=4) of the Brillouin zone for PbTe are nonparabolic[21, 24, 30, 31] and can be well described by a Kane band model.

Furthermore, the scattering of charge carriers in PbTe is known to be dominated by acoustic phonons[21, 25] in the temperature and carrier concentration range having high thermoelectric performance, as is the case for most good thermoelectric material. This is demonstrated in Fig 2a showing the hall mobility decreases sharply with temperature ($\mu \sim T^p$ where $p < -1.5$[21]). Other scattering mechanisms such as by grain boundaries, polar-optical phonons, ionized impurities, predict $p \geq -\frac{1}{2}$ implying that these mechanisms do not dominate the transport properties. In fact, the Hall mobility predicted by single Kane band model (SKB), with acoustic scattering theory[18, 19] and temperature dependent $m^*$ (Fig 2b), agrees well with the experimental data (Fig. 2a), for both La- and I-doped PbTe.

Such a Kane band model provides the expressions for the transport coefficients[21, 24] as follows:
Hall carrier density

$$n_H = \frac{1}{eR_H} = A^{-1} \frac{N_v(2m_b^*k_BT)^{3/2}}{3\pi^2\hbar^3} {}^0F_0^{3/2} \quad (1);$$

Hall factor

$$A = \frac{3K(K+2)}{(2K+1)^2} \frac{{}^0F_{-4}^{1/2} \cdot {}^0F_0^{3/2}}{({}^0F_{-2}^1)^2} \quad (2);$$

Hall mobility

$$\mu_H = A \frac{2\pi\hbar^4 eC_l}{m_c^*(2m_b^*k_BT)^{3/2}E_{def}^2} \frac{3{}^0F_{-2}^1}{{}^0F_0^{3/2}} \quad (3);$$

Seebeck coefficient

$$S = \frac{k_B}{e}\left[\frac{{}^1F_{-2}^1}{{}^0F_{-2}^1} - \xi\right] \quad (4);$$

and Lorenz number

$$L = \left(\frac{k_B}{e}\right)^2 \left[\frac{{}^2F_{-2}^1}{{}^0F_{-2}^1} - \left(\frac{{}^1F_{-2}^1}{{}^0F_{-2}^1}\right)^2\right] \quad (5);$$

where ${}^nF_k^m$ has a similar form as the Fermi integral

$${}^nF_k^m = \int_0^\infty \left(-\frac{\partial f}{\partial \varepsilon}\right)\varepsilon^n(\varepsilon + \alpha\varepsilon^2)^m[(1+2\alpha\varepsilon)^2 + 2]^{k/2} d\varepsilon \quad (6).$$

In the above equations, $k_B$ is the Boltzmann constant, $\hbar$ the Boltzmann constant, $C_l$ the combined elastic moduli[19], $E_{def}$ the deformation potential coefficient[19] characterizing the strength of carriers scattered by acoustic phonons, $\xi$ the reduced Fermi level, $\varepsilon$ the reduced energy of the electron state, $\alpha$ ($=k_BT/E_g$) the reciprocal reduced band separation ($E_g$, at L point of the Brillouin zone in this study) and $f$ the Fermi distribution. This model also considers an ellipsoidal Fermi surface by taking the ratio of the longitudinal ($m_\parallel^*$) to transverse ($m_\perp^*$) effective mass components into account via the term $K = m_\parallel^* / m_\perp^*$.

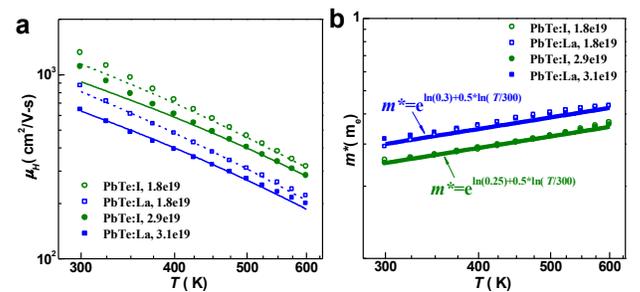

Fig. 2. Temperature dependent Hall mobility (a) and effective mass (b) for La- and I-doped PbTe. The experimental Hall mobility (symbols) can be well predicted (curves) by an acoustic scattering mechanism. La-doping leads to a ~20% higher effective mass over the entire temperature range. The increase in effective mass with increasing temperature is due to the Kane type band structure and is associated with the temperature dependent band gap.

Utilizing the above SKB model, excellent prediction of the Hall carrier concentration dependent Seebeck coefficient and Hall mobility can be obtained for both La- and I-doped PbTe over a broad carrier concentration range as shown in Fig. 3. Literature data from different sources[27, 32-38] show good consistency with the current work.

It is seen that the La-doped series shows slightly higher Seebeck coefficient values at both 300 K (Fig. 3a) and 600 K (Fig. 3b), which correspondingly means a higher density-of-states effective mass by 20% than that in the I-doped samples. Quantitatively, $m^*$ is found to be $0.25\pm0.03$ $m_e$ and $0.30\pm0.02$ $m_e$ at 300 K, and $0.35\pm0.02$ $m_e$ and $0.41\pm0.02$ $m_e$ at 600 K, for I- and La-doped PbTe, respectively, where the standard deviations are obtained on approximately 10 different samples. Most importantly, only varying $m^*$ by 20%, enables an accurate prediction (curves in Fig. 3c and 3d) of the Hall mobility at both 300 and 600 K using the SKB model without any other adjustable parameters. Here, the values $K=3.6$[39], $C_l=7.1\times10^{10}$ Pa[25], $E_{def}=22$ eV[27] and $\alpha=k_BT/(0.18\text{ eV} + 0.0004\text{ eV/K} \times T)$ [21, 40-42] are used for both I- and La-doped series. The excellent agreement between the experimental and predicted results confirms the validity of the model itself and additionally indicates that the higher $m^*$ is indeed responsible for the observed higher $S$ and lower $\mu_H$.

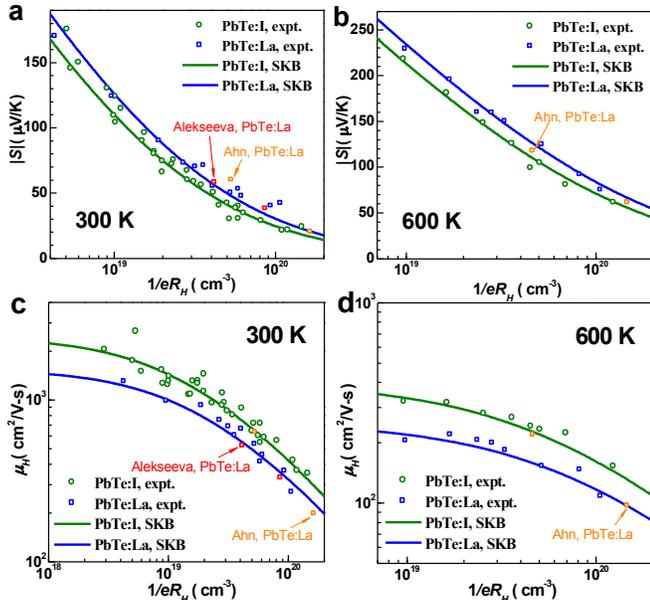

Fig. 3. Hall carrier concentration versus Seebeck coefficient (a and b) and Hall mobility (c and d) for I- and La-doped PbTe at 300 K (a and c) and 600 K (b and d), compared with the predicted results (curves) according to the single Kane band model. Providing a 20% higher effective mass in La-doped series, both the increase in Seebeck coefficient and decrease in Hall mobility can be well predicted by the SKB model.

The higher $m^*$ in La-doped PbTe is presumably due to the conduction band flattening, related to an increase in band gap[32] according to the Kane dispersion $E(k)$ [24, 43]:

$$\frac{\hbar k^2}{2m^*} = E\left(1 + \frac{E}{E_g}\right) \quad (7).$$

In a Kane band system, the increase of $m^*$ with increasing band gap has been theoretically predicted [21, 44] and experimentally confirmed [31, 39, 45-47] in lead chalcogenides. Furthermore, the increase of $m^*$ in PbTe and related materials can be induced by either temperature[21, 39, 45] or chemical substitution[15], making available an additional tunable parameter for further investigation of $m^*$ dependent thermoelectric properties.

With the knowledge of band separation at L point of the Brillouin zone, one can calculate the reduced Fermi level from the experimental Seebeck coefficient according to Eq. 4. Consequently, $m^*$ can be obtained from Eqs. 1-2. In this way, we calculate the temperature dependent $m^*$ (Fig. 2b) for both La- and I-doped PbTe having room temperature Hall carrier concentrations of ~1.8 and ~$3\times10^{19}$ cm$^{-3}$. It is clearly seen that the 20% higher $m^*$ in the La-doped series persists throughout the entire temperature range. The $m^*$, obtained in the manner is related to the density-of-states at the band edge. Largely resulting from the lattice expansion[21], the band gap increases with increasing temperature leading to an increase in $m^*$ as theoretically predicted [21, 44] and experimentally observed [31, 39, 45-47] in Kane band systems (Eq. 7). Therefore, the observed increase in $m^*$ with increasing temperature by $d\ln m^*/d\ln T = 0.5$ (curves), can be well understood by the SKB model and consistent with literature reports[21, 39, 45].

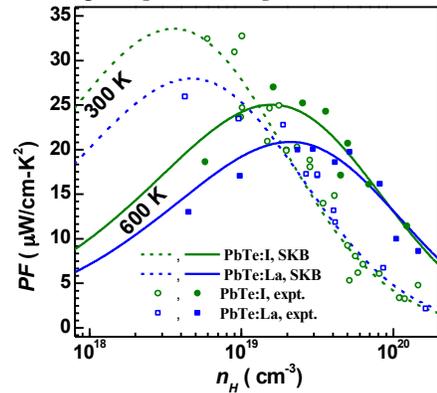

Fig. 4. Thermoelectric power factor versus Hall carrier concentration for La- and I-doped PbTe at the two end temperatures of 300 and 600 K. The 20% difference in effective mass due to variant dopant leads to the peak power factor differing by ~20% at both temperatures. The ~40% increase in effective mass (Fig. 2b), originating from the temperature increase from 300 to 600 K, results in a ~35% decrease in peak power factor in both La- and I-doped samples.

With a combination of the predicted Hall carrier concentration dependent Seebeck and Hall mobility, the thermoelectric power factor ($PF= S^2n_He\mu_H$) is calculated and compared with the experimental data for the La- and I-doped series at 300 and 600 K in Fig. 4. It is now clear that ~20% higher $m^*$ leads to ~20% lower maximal $PF$ in La-doped series at both 300 K (34 vs. 28 μW/cm-K$^2$) and 600 K (25 vs. 21 μW/cm-K$^2$). Moreover, the temperature induced ~40% increase in $m^*$ correspondingly results in a ~35% decrease in maximal $PF$ for both La- (28 vs. 21 μW/cm-K$^2$) and I-doped (34 vs. 25 μW/cm-K$^2$) PbTe, as temperature rises from 300 to 600 K. An increase of $m^*$ due to independent mechanisms leads to a reduction of the overall optimal thermoelectric power factor, despite the resulting increase of the Seebeck coefficient.

The physics behind why a higher $m^*$ without increasing $N_v$ has a detrimental effect to thermoelectric performance, is similar for the SKB model as it is for a parabolic band model. Combining Eqs. 1, 3 and 4, one obtains:

$$PF = \frac{2N_v \hbar k_B^2 C_l}{\pi E_{def}^2} \cdot \frac{1}{m_c^*} \cdot \left(\frac{{}^1F_{-2}^1}{{}^0F_{-2}^1} - \xi\right)^2 {}^0F_{-2}^1 \qquad (8).$$

Because the first term only includes fundamental constants or composition independent material parameters in this study, the $PF$ is inversely proportional to $m_c$ which is proportional to $m^*$ in cubic materials. The third term is a function of the reduced Fermi level and reduced band separation. This term has the same maximal value when tuning the doping level at a given reduced band separation (meaning a given temperature in this study). Therefore, a temperature induced increase of $m^*$ in a Kane band system may lead to a slight deviation from the relationship of $PF \propto 1/m_c^*$ as predicted in a parabolic band system where the third term is not band separation dependent.

With the known temperature dependent $m^*$ for both La- and I-doped PbTe, use of the SKB model also enables a prediction of the temperature dependent transport properties at any temperature and doping level. Fig. 1 also shows the calculated results (curves) having the same doping level as the actual samples, using the experimentally estimated lattice thermal conductivity. Possessing a comparable thermal conductivity at similar doping levels (Fig. 1c) the La-doped series shows ~20% lower $zT$ due to the ~20% higher $m^*$ over the entire temperature range studied here, even though the Seebeck coefficient is higher, as compared to the I-doped samples.

In summary, we show an example of achieving higher thermoelectric performance as a result of lower effective mass, which is contrary to the generally held belief that higher effective mass is beneficial for thermoelectrics because of the resulting higher Seebeck coefficient. It is demonstrated that the significant reduction of carrier mobility resulting from increased effective mass through band flattening actually reduces the thermoelectric power factor in n-PbTe and $zT$. Our efforts here have shown that light band mass leads to higher performance and should be used as an important strategy for discovering and improving thermoelectric materials.

This work is supported by NASA-JPL and DARPA Nano Materials Program.